\documentclass[pra,twocolumn,nofootinbib,floatfix,10pt]{revtex4-2}
\usepackage{textcomp,mathcomp}
\usepackage{amsmath}
\usepackage{amssymb}
\usepackage{wasysym}
\usepackage{graphicx}
\usepackage{color,soul}
\usepackage{physics}
\usepackage{siunitx}
\usepackage{dsfont}
\usepackage{float}
\usepackage[english]{babel}
\usepackage{blindtext}
\usepackage[english,nomargin,inline,marginclue,draft]{fixme}
\pdfpageheight\paperheight
\pdfpagewidth\paperwidth

\usepackage[colorlinks,linkcolor=blue,anchorcolor=blue,citecolor=blue,urlcolor=blue]{hyperref}
\usepackage{balance}

\fxusetheme{colorsig}
\FXRegisterAuthor{cg}{acg}{CG}  
\FXRegisterAuthor{th}{ath}{\color{blue}TH}  
\FXRegisterAuthor{ib}{aib}{\color{red}IB} 
\FXRegisterAuthor{sh}{ash}{\color{cyan}SH} 
\FXRegisterAuthor{db}{adb}{\color{green}DB} 
\FXRegisterAuthor{ps}{aps}{PS}
\makeatletter
\renewcommand*\FXLayoutInline[3]{%
  {\@fxuseface{inline}\ignorespaces{\color{fx#1}[#3: #2]}}}
\makeatother

\long\def\symbolfootnote[#1]#2{\begingroup%
\def\thefootnote{\fnsymbol{footnote}}\footnotetext[#1]{#2}\endgroup}

\def\nobreakbefore{%
  \relax\ifvmode\else
    \ifhmode
      \ifdim\lastskip > 0pt\relax
        \unskip\nobreakspace
      \else 
        \nobreakspace
      \fi
    \fi
  \fi
}
\let\oldcite\cite
\renewcommand\cite{\nobreakbefore\oldcite}

\begin{document}
\title{Observation of a Rydberg-atom time crystal with an ultralong lifetime}

\author{Qi-Feng Wang$^{1,2,\textcolor{blue}{\star}}$}
\author{Tian-Yu Han$^{1,2,\textcolor{blue}{\star}}$}
\author{Ya-Jun Wang$^{1,2,\textcolor{blue}{\star}}$}
\author{Dong-Yang Zhu$^{1,2,\textcolor{blue}{\star}}$}
\author{Chao Yu$^{1,2}$}
\author{Yu Ma$^{1,2}$}
\author{Yi-Ming Yin$^{1,2}$}
\author{Guang-Can Guo$^{1,2}$}
\author{Bang Liu$^{1,2,\textcolor{blue}{\ddagger}}$}
\author{Li-Hua Zhang$^{1,2,\textcolor{blue}{\#}}$}
\author{Dong-Sheng Ding$^{1,2,\textcolor{blue}{\dagger}}$}
\author{Bao-Sen Shi$^{1,2}$}

\affiliation{$^1$Laboratory of Quantum Information, University of Science and Technology of China, Hefei, Anhui 230026, China.}
\affiliation{$^2$Anhui Province Key Laboratory of Quantum Network, University of Science and Technology of China, Hefei 230026, China.}

\date{\today}
\symbolfootnote[1]{Q.-F.W., T.-Y.H., Y.-J.W., and D.-Y.Z. contributed equally to this work.}
\symbolfootnote[4]{lb2016wu@ustc.edu.cn}
\symbolfootnote[3]{zlhphys@ustc.edu.cn}
\symbolfootnote[2]{dds@ustc.edu.cn}

\maketitle

\textbf{Continuous time crystals (CTCs) represent a nonequilibrium quantum phase that spontaneously breaks time-translation symmetry without periodic external driving, manifesting as persistent, long-lived oscillations under steady pumping. The lifetime is constrained by the instability of the limit cycle phase, balanced between nonlinear feedback and energy dissipation, which have rarely been studied in experiments before. Here, we report an observation of an ultralong-lived Rydberg-atom CTC in a driven-dissipative many-body atomic system. By harnessing long-range interactions and engineering a dissipative environment that stabilizes the limit-cycle dynamics, we suppress heating and decay effects that typically destroy time-crystalline order. The key factor underlying the ultralong-lived CTC is the closing of the Liouvillian gap and the near-zero real part of the system's Liouvillian eigenspectrum. Through systematic optimization, we achieve an oscillatory lifetime exceeding 16.95 hours—orders of magnitude longer than previous CTC realizations. Our work establishes a robust platform for exploring long-lived autonomous nonequilibrium phases and paves the way for applications in quantum sensing and continuous-time quantum information processing.}

\section*{INTRODUCTION}

Achieving long-lived many-body states in the nonequilibrium dynamics of interacting quantum systems remains a fundamental challenge \cite{serbyn2021quantum,defenu2024out,ho2023quantum}. In isolated or weakly dissipative systems, unitary evolution typically drives the system toward thermal equilibrium, which is governed by the eigenstate thermalization hypothesis (ETH) \cite{srednicki1994chaos} and rooted in Boltzmann's ergodic hypothesis \cite{boltzmann1887ueber}. To sustain a long-lived many-body state, it is therefore essential to suppress thermalization and decay over extended timescales. A paradigmatic example is the strongly interacting Rydberg atom array \cite{bernien2017probing,turner2018weak,browaeys_many-body_2020, bluvstein_controlling_2021}, where nonergodic dynamics unfold within a constrained sub-Hilbert space, giving rise to coherent revivals of the $Z_2$ state with a long lifetime. In strongly dissipative systems such as continuous time crystals (CTCs) \cite{wilczek2012quantum,kongkhambut2022observation} and quantum synchronization \cite{PhysRevLett.88.230602,PhysRevLett.93.204103,PhysRevLett.98.184101}, long lifetime arises from a stable limit cycle attractor \cite{buvca2019non,ding2023ergodicity,dutta2025quantum}. On such an attractor, transverse relaxation vanishes in the thermodynamic limit, allowing fluctuations to decay over infinitely long times, while undamped oscillatory motion along the cycle sustains persistent periodicity. Therefore, extending the long lifetime of time crystals requires precise control of the interplay among interactions and dissipation, for instance, by tuning the mean-field excitation number to suppress phase diffusion \cite{cabot2024nonequilibrium}.

\begin{figure*}
    \centering
    \includegraphics[width=1\linewidth]{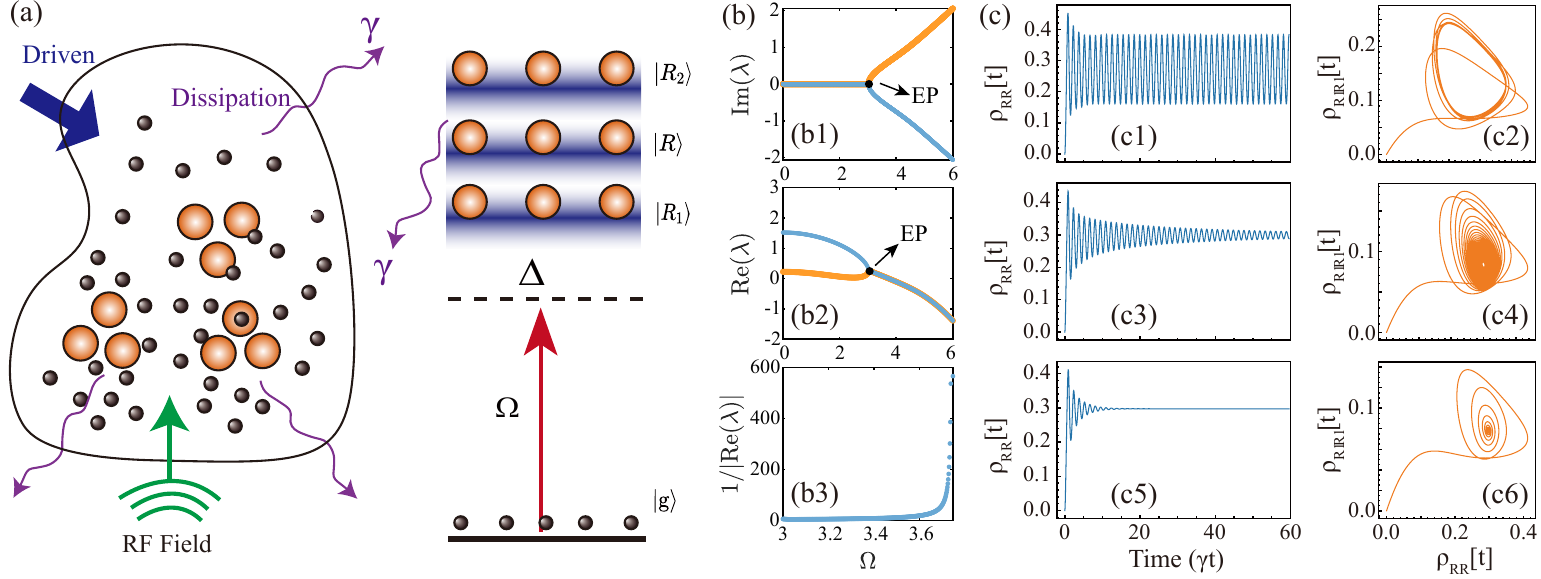}
    \caption{\textbf{Diagram of the many-body system and simulations.} (a) Schematic diagram of the drive-dissipation system, incorporating laser and RF field driving as well as the system's intrinsic dissipation. The simplified energy diagram of the system consists of a ground state $\ket{g}$ and three Rydberg states $\ket{R},\ket{R_1}$ and $\ket{R_2}$ dressed by RF-field. $\Omega$ is the Rabi frequency coupling the ground state and Rydberg state with the laser, $\Delta$ is the detuning between the laser and the atomic resonant transition, and $\gamma$ is the decay rate of the Rydberg energy levels. (b) Near zero frequency, the simulated imaginary part of the Liouvillian eigenspectrum scales as (b1), the real part as (b2), and the critical behavior of the reciprocal of the real part near zero scales as (b3). The black dot marks the Liouvillian exceptional point (EP). The parameters are set as $\{\Omega_1, \Omega_2, \Gamma_{\rm{eff}},\Gamma_{\rm{eff1}},  \Gamma_{\rm{eff2}},\gamma,\gamma_1,\gamma_2\}=\{3.5,1.6,11,7,1.15,2.9,3.2,2.9\}$ and the energy detuning is $\Delta=\Delta_1=\Delta_2=0$. (c) System dynamics evolution under different decay rates, where (c1), (c3), and (c5) correspond to $\gamma=1$, 1.3, and 1.7, respectively, and (c2), (c4), and (c6) correspond to evolutionary trajectories in the phase plane.}
    \label{fig1}
\end{figure*}

The lifetimes of CTCs vary widely across different platforms, ranging from approximately milliseconds in ultracold atom-cavity systems \cite{kongkhambut2022observation}, thermal Rydberg gases \cite{wu2024dissipative,liu2024bifurcation,jiao2025observation}, and erbium-doped solid-state systems \cite{chen2023realization}, to over 40 minutes in electron-nuclear spin systems \cite{greilich2024robust}. To achieve long lifetime, one needs to close the real-part gap in the Liouvillian energy spectrum and isolate the time-crystalline modes from fast-decaying ones \cite{iemini2018boundary}. The strong interactions in Rydberg atoms \cite{saffman2010quantum,adams2019Rydberg,browaeys_many-body_2020}, combined with controlled dissipation, make them a powerful platform for studying diverse nonequilibrium phenomena, such as self-organization and nonequilibrium phase transitions \cite{lee2012collective,carr2013nonequilibrium,helmrich2020signatures,ding2019Phase,ding2022enhanced,wadenpfuhl2023emergence,ding2023ergodicity}, time crystals \cite{wu2024dissipative,liu2024bifurcation, liu2024higher, jiao2025observation}, and time quasicrystals \cite{zhu2025observation}. However, the lifetimes observed in CTCs have typically been limited to the order of milliseconds in Rydberg atom systems. Breaking through the barrier of short lifetime to achieve ultralong-lived time crystals \cite{machado2020long,zaletel2023colloquium} would not only validate the existence of stable nonequilibrium phases in the thermodynamic limit but also provide a powerful testbed for exploring the interplay between dissipation, interactions, and quantum correlations over macroscopic timescales \cite{cabot2024nonequilibrium}. In addition, a time crystal with a lifetime orders of magnitude longer than its microscopic timescales may serve as a robust and sensitive quantum sensor for external fields, or a memory element for quantum information processing \cite{sacha2018time}. Nevertheless, a detailed and in-depth investigation of ultralong-lived time crystals in Rydberg atom systems has yet to be reported. 

In this work, we have observed an ultralong lifetime CTC in a driven-dissipative Rydberg atom ensemble with a lifetime exceeding 16.95 hours—an improvement of several orders of magnitude compared to previous realizations. By harnessing the strong long-range Rydberg interactions and engineering a dissipative environment through controlled optical pumping and external electric fields, we stabilize the limit-cycle dynamics that underpin the time-crystalline order. We identify that the key factor governing the CTC long lifetime is the near-zero real part of the Liouvillian eigenspectrum. Through careful optimization of these parameters, we achieve a persistent oscillatory state in which transverse relaxation is suppressed, allowing the system to realize a limit-cycle attractor with sub-Hz frequency stability over macroscopic timescales. This work establishes a robust and highly tunable platform for exploring long-lived autonomous nonequilibrium phases in Rydberg atom systems and opens new avenues for practical applications in quantum sensing, precision metrology, and continuous-time quantum information processing.

\begin{figure*}
\centering
\includegraphics[width=0.93\linewidth]{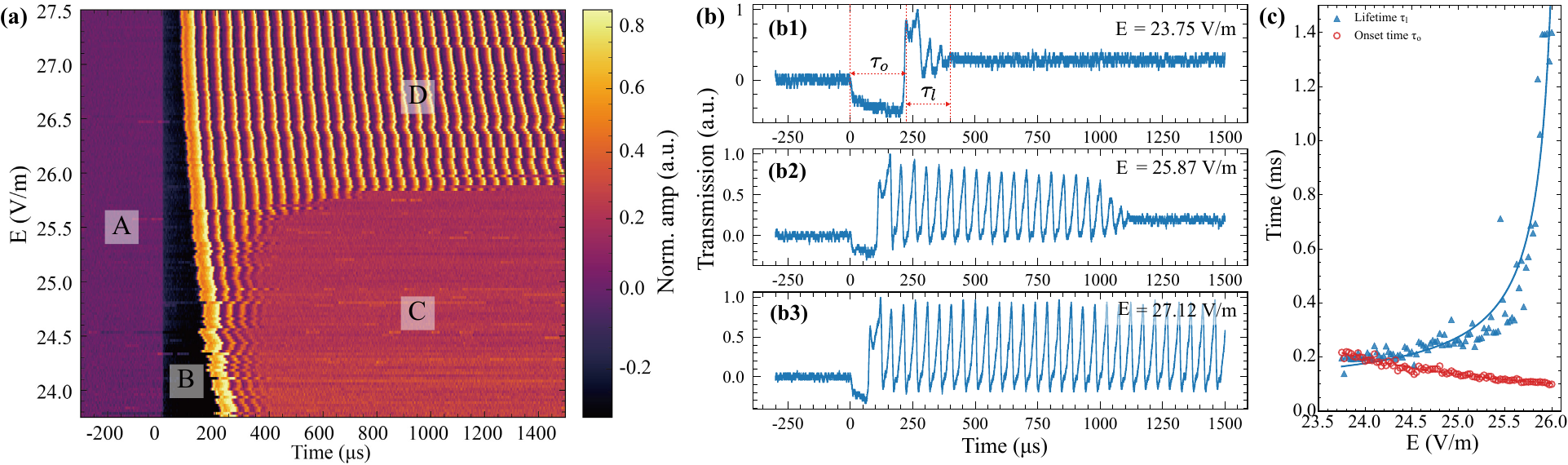}
\caption{\textbf{Measured CTC criticality and lifetime versus the RF-field amplitude.} (a) Measured transmission after a step driving field is suddenly switched on at $t=0$. A uniform field $E = V_{\mathrm{pp}}/d$ is applied across two parallel plates separated by $d = 4$~cm, and its amplitude is scanned over $E = 23.75$--$27.5$~V/m; the color scale shows the normalized transmission amplitude. Four regimes are labeled: (A) before the field is applied ($t<0$), where no oscillation occurs; (B) the field is on but the time crystal has not yet formed, defining the onset time $\tau_{\mathrm{o}}$; (C) the time crystal has formed with a finite lifetime $\tau_{\mathrm{l}}$ that increases with $E$; and (D) the time crystal persists throughout the record, its apparent lifetime being limited only by truncation of the $1.5$~ms acquisition window. (b) Normalized transmission time traces extracted from (a) at $E = 23.75$~V/m (b1), $25.87$~V/m (b2) and $27.12$~V/m (b3); the onset time $\tau_{\mathrm{o}}$ and lifetime $\tau_{\mathrm{l}}$ are marked in (b1). The oscillation is short-lived in (b1), persists for a finite time before decaying in (b2), and shows no decay within the measurement window in (b3). (c) Onset time $\tau_{\mathrm{o}}$ (red open circles) and lifetime $\tau_{\mathrm{l}}$ (blue triangles) as functions of the driving field $E$. With increasing field the onset time decreases slightly, whereas the lifetime grows rapidly toward the high $E$ end of the scan. The lifetime is fitted (solid blue curve) to $\tau_{\mathrm{l}}(E) = a_{0}/(E_{\mathrm{c}} - E) + \tau_{\mathrm{off},0}$, and the onset time is fitted (dashed red curve) to $\tau_{\mathrm{o}}(E) = a_{1}\,(E - E_{\mathrm{c}}')^{\alpha_{1}}$, where the fit parameters are given in the main text. Higher $E$ points are excluded because their lifetimes exceed the $1.5$~ms measurement window.}
\label{fig2}
\end{figure*}

\section*{Physical model}
To capture the emergence of CTC behavior in a driven Rydberg atomic system, we model the experiment as a nonequilibrium ensemble of interacting atoms subject to coherent driving and dissipation. Rydberg excitations in the ensemble experience strong van der Waals interactions characterized by $V = C_6 / r^6$, which introduce intrinsic nonlinearity into the many-body dynamics. An external radio-frequency (RF) electric field with amplitude $E_{\rm RF}$ and frequency $\omega$ is applied to the system, predominantly modulating the Rydberg level while leaving the lower-lying states essentially unaffected, as shown in Fig.~\ref{fig1}(a). The RF field dresses the Rydberg state and induces Floquet sidebands in the Rydberg excitation spectrum. To describe the physics, we construct an effective driven-dissipative model in which each atom is reduced to a ground state $\ket{g}$ coupled to a set of RF-dressed Rydberg states, including the bare state $\ket{R}$ and the dominant $\pm1$-order sidebands $\ket{R_{1,2}}$, with corresponding effective Rabi frequencies $\Omega_{1,2}$, detuning $\Delta_{1,2}$, and a decay rate $\gamma$, the energy level diagram is displayed in Fig.~\ref{fig1}(a). The Hamiltonian is:
\begin{equation}
\begin{aligned}
    \hat{H} & =\frac{1}{2}\sum_{i}\left(\Omega\sigma_{i}^{gR}+\Omega_{1}\sigma_{i}^{gR_1}+\Omega_{2}\sigma_{i}^{gR_2}+h.c.\right)\\ & -\sum_{i}\left(\Delta n_{i}^{R}+\Delta_{1}n_{i}^{R_1}+\Delta_{2}n_{i}^{R_2}\right)\\ & +\sum_{i\neq j}\bigg[V_{ij}^{RR_1}n_{i}^{R}n_{j}^{R_1}+V_{ij}^{RR_2}n_{i}^{R}n_{j}^{R_2}+V_{ij}^{R_1R_2}n_{i}^{R_1}n_{j}^{R_2}\\ &+\frac{1}{2}(V_{ij}^{RR}n_{i}^{R}n_{j}^{R}+V_{ij}^{R_1R_1}n_{i}^{R_1}n_{j}^{R_1}+V_{ij}^{R_2R_2}n_{i}^{R_2}n_{j}^{R_2})\bigg]
\end{aligned}
\end{equation}
where $\sigma_{i}^{gr}$ ($r={R,R_1,R_2}$) represents the $i$-th atom transition between the ground state $\left| g \right\rangle$ and the Rydberg state $\left|  r \right\rangle$, $n_{i}^{R,R_1,R_2}$ are the population operators of the Rydberg energy levels $\left|  R \right\rangle$ and $\left|  R_1 \right\rangle$, and $\left|  R_2 \right\rangle$, and $V_{ij}^{RR_1}$, $V_{ij}^{RR_2}$, and $V_{ij}^{R_1R_2}$ are the interactions between the Rydberg atoms in states $\left|R\right\rangle$, $\left|R_{1}\right\rangle$, and $\left|R_{2}\right\rangle$, respectively. Under the mean-field treatment, the energy level shift and dissipation induced by the interaction between atoms in each Rydberg state can be equivalent by an effective detuning term ($\Delta-i\gamma_{\rm{eff}}$, $\Delta_1-i\gamma_{\rm{eff1}}$, and $\Delta_2-i\gamma_{\rm{eff2}}$). Within this framework, the competition between nonlinear interactions and continuous energy exchange with the environment drives the system away from a stationary state into a stable limit-cycle phase.

Theoretically, the dynamical properties of this system are described by the Lindblad master equation
\begin{equation}
\begin{aligned}
\dot{\rho}=-i[H, \rho]+\sum_j\left(L_j \rho L_j^{\dagger}-\frac{1}{2}\left\{L_j^{\dagger} L_j, \rho\right\}\right):=\mathcal{L}\rho,
\end{aligned}
\end{equation}
where $\mathcal{L}$ is a Liouvillian superoperator and the Lindblad jump operators take the form $L_1 = \sqrt{\gamma_1}|g\rangle\langle R_1|$, $L_2 = \sqrt{\gamma}|g\rangle\langle R|$, and $L_3 = \sqrt{\gamma_2}|g\rangle\langle R_2|$. The Liouvillian eigenspectrum is obtained by solving $\mathcal{L}\hat{\rho}_n = \lambda_n \hat{\rho}_n$, where $\lambda_n$ and $\hat{\rho}_n$ denote the Liouvillian eigenvalues and eigenstates, respectively. Figure~\ref{fig1}(b) illustrates the Liouvillian spectral characteristics, where the real part represents the system's dissipative properties and the imaginary part denotes the oscillation frequency of the density matrix states. By analyzing the Liouvillian eigenspectrum, we obtain the oscillation and dissipation characteristics of the density matrix state. We observe that the `gap-closing' of the real part drives the system across an exceptional point (EP), transitioning from a steady state ($\rho(t)=\rho e^{\lambda t}$, $\rm{Im(\lambda)=0}$) to a non-zero oscillatory regime [Figs.~\ref{fig1}(b1)-(b2)]. In the oscillatory regime, a critical dissipationless oscillatory state ($\rho(t)=\rho e^{i\rm{Im(\lambda)}t}$, $\rm{Re(\lambda)=0}$) appears as the real part of the Liouvillian eigenspectrum approaches zero~\cite{buvca2022algebraic,cabot2024nonequilibrium}. 

In general, the lifetime is inversely proportional to dissipation: higher dissipation leads to a shorter lifetime. Figure~\ref{fig1}(b3) shows the relationship between the inverse of the real part of the eigenspectrum (representing dissipation) and the parameter $\Omega$, which is experimentally tunable via the driving electric field strength that controls the coupling to Rydberg sidebands. As the decay rate approaches zero, the oscillatory state achieves an ultralong lifetime, corresponding to a long-lived time crystal phase. Furthermore, Fig.~\ref{fig1}(c) shows the time evolution of density matrix $\rho_{RR}$ and its phase-space trace under different parameter conditions. Long-lived coherent oscillations with negligible damping appear in Figs.~\ref{fig1}(c1)-(c2), while in Figs.~\ref{fig1}(c3)-(c6), the oscillations in \(\rho_{RR}\) are damped and the system relaxes toward a nonoscillatory state.

\begin{figure*}
\centering
\includegraphics[width=1\linewidth]{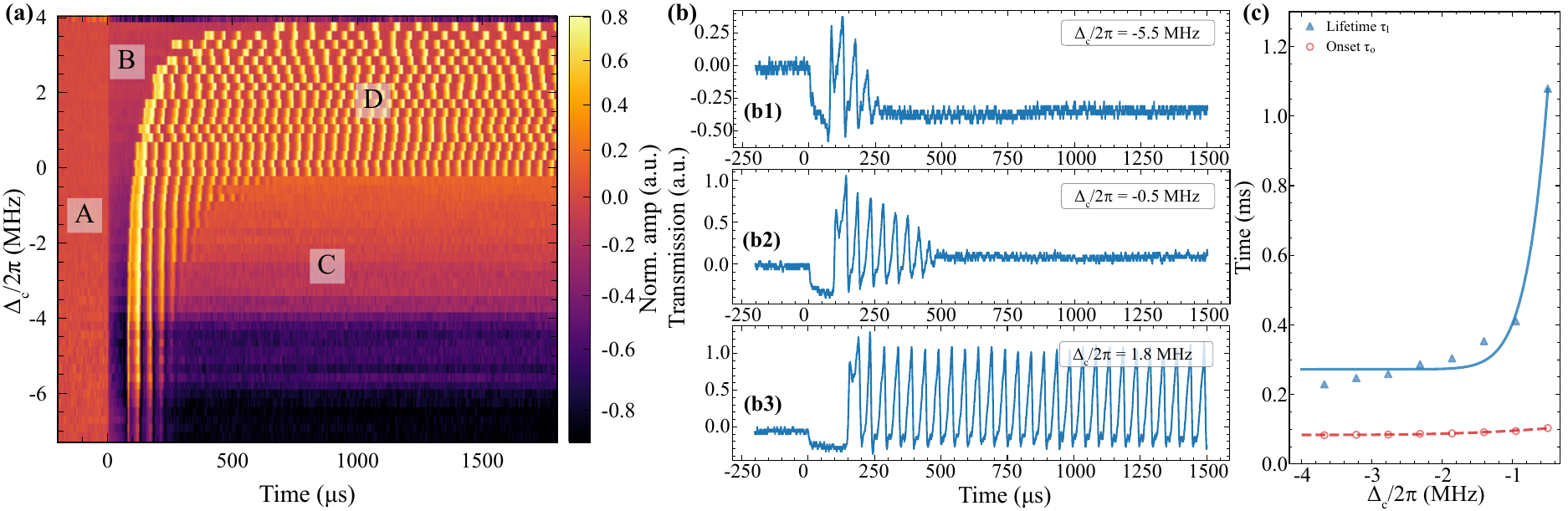}
\caption{\textbf{Measured criticality and lifetime of CTC versus the laser detuning.} (a) Measured transmission dynamics under the driving of a step RF-field with $E = 26.5$~V/m at $t = 0$. The laser detuning $\Delta_c$ is scanned over $-2\pi \times 7.3$ to $2\pi \times 4.0$~MHz; the color scale represents the normalized transmission amplitude. The system exhibits four regimes: no time crystal (A), onset regime (B), fast-decay regime (C), long-lived regime (D). (b)~Normalized transmission time traces extracted from (a) at $\Delta_c = -2\pi \times 5.5$~MHz (b1), $-2\pi \times 0.5$~MHz (b2), and $2\pi \times 1.8$~MHz (b3). The oscillation is short-lived at red detuning (b1), persists for a finite time before decaying near resonance (b2), and exhibits long-lived behavior (b3). (c)~Measured onset time $\tau_{\mathrm{o}}$ (red open circles) and lifetime $\tau_{\mathrm{l}}$ (blue triangles) as functions of the laser detuning $\Delta_c$. With increasing detuning toward resonance, both the onset time and the lifetime increase. Within the fitted region ($-4.0 \leq \Delta_c/2\pi \leq -0.5$~MHz), the lifetime exhibits a sharp power-law rise: $\tau_{\mathrm{l}}(\Delta_c) = \tau_{\mathrm{off},2} + a_2(\Delta_c/2\pi - \Delta_{\mathrm{min}}/2\pi)^{\alpha_{2}}$ with $\alpha_{2} \approx 11.7$, and the onset time follows $\tau_{\mathrm{o}}(\Delta_c) = \tau_{\mathrm{off},3} + a_3(\Delta_c/2\pi - \Delta_{\mathrm{min}}/2\pi)^{\alpha_{3}}$ with $\alpha_{3} \approx 3.0$, where $\Delta_{\mathrm{min}}/2\pi = -4.0$~MHz is the lower boundary of the fit range. The fit parameters are given in the main text.}
\label{fig3}
\end{figure*}

\section*{Results}

\subsection{Measured criticality and CTC lifetime}

To measure the CTC lifetime, we systematically investigated the dependence behavior on the amplitude of an applied electric field. In the experiment, we recorded the temporal dynamics of probe transmission upon switching on the electric field at different amplitudes $E$. This approach allowed us to identify the critical point at which the system transitions from the thermal phase to the limit cycle phase, corresponding to the `gap-closing' of the real part of Liouvillian eigenspectrum. The resulting heatmap is shown in Fig.~\ref{fig2}(a). A clear trend emerges: increasing the amplitude $E$ leads to more observable oscillation cycles, indicating enhanced stability of the time-crystalline response at larger $E$. For instance, at $E$ = 23.75 V/m (Fig.~\ref{fig2}(b1)), the system exhibits several oscillation cycles that decay rapidly. In contrast, at $E$ = 25.87 V/m (Fig.~\ref{fig2}(b2)) and $E$ = 27.12 V/m (Fig.~\ref{fig2}(b3)), the system shows a significantly longer persistent oscillation time $\tau_{\mathrm{l}}$, which corresponds to the CTC lifetime. The limited lifetime of CTCs observed in Figs.~\ref{fig2}(b1-b2) suggests that the nonzero real part of the Liouvillian eigenspectrum gives rise to strong dissipation, even though the `gap-closing' effect emerges.

Beyond the lifetime itself, we find that the CTC lifetime $\tau_{\mathrm{l}}$ depends on the onset time $\tau_{\mathrm{o}}$, i.e., the time required for the system to evolve from the initial state to the limit cycle attractor. A longer oscillation time $\tau_{\mathrm{l}}$ is consistently accompanied by a shorter onset time $\tau_{\mathrm{o}}$, suggesting that fields which rapidly drive the system into the limit cycle also yield a more robust CTC. We record the relationship among the CTC onset time $\tau_{\mathrm{o}}$, the lifetime $\tau_{\mathrm{l}}$, and the electric field amplitude $E$, as presented in Fig.~\ref{fig2}(c). The measured lifetime is well fitted to $\tau_{\mathrm{l}}(E) = a_{0}/(E_{\mathrm{c}} - E) + \tau_{\mathrm{off},0}$ with $a_{0} = 0.243$~ms$\cdot$V/m, $E_{\mathrm{c}} = 26.2$~V/m and $\tau_{\mathrm{off},0} = 0.064$~ms (coefficient of determination, $R^2 = 0.90$). The onset time is fitted to a power law function $\tau_{\mathrm{o}}(E) = a_{1}\,(E - E_{\mathrm{c}}')^{\alpha_{1}}$ with $a_{1} = 0.904$~ms$\cdot$(V/m)$^{-\alpha_{1}}$, $E_{\mathrm{c}}' = 20.8$~V/m and $\alpha_{1} = -1.32$ ($R^2 = 0.95$). The results indicate that increasing $E$ not only extends $\tau_{\mathrm{l}}$ but also reduces $\tau_{\mathrm{o}}$, highlighting a field-strength-dependent correlation between rapid attractor acquisition and prolonged oscillatory persistence. The electric field reshapes the Liouvillian energy spectrum landscape. A larger electric field raises the potential barrier that separates the CTC phase from the thermal phase. This elevated barrier suppresses thermal fluctuations and slows down the system's escape from the time-crystalline attractor, thereby prolonging the CTC lifetime.

\begin{figure*}
\centering
\includegraphics[width=1\linewidth]{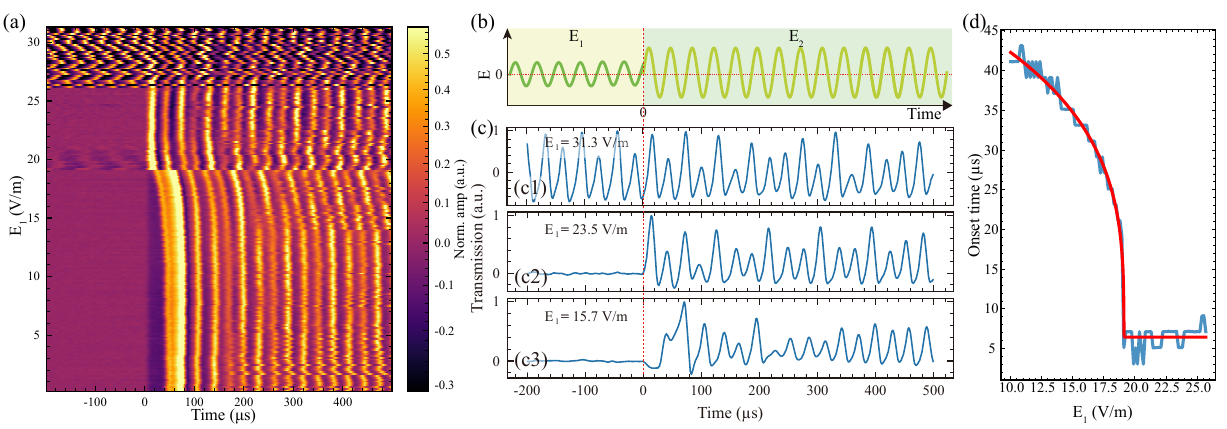}
\caption{\textbf{Initial-state-dependent onset dynamics.} (a) Probe transmission versus time as the amplitude of the RF-field $E_1$ is scanned from $0$ to $31.3$~V/m for $t\leq0$ and $E_2 = 31.3$~V/m for $t>0$; the color scale gives the normalized transmission amplitude. The different values of $E_1$ affect the evolution of the system at $t>0$. (b) Schematic of the two-stage driving protocol: $E_1$ prepares the initial state for $t>0$ and is switched at $t = 0$ to a fixed amplitude $E_2 = 31.3$~V/m with the same frequency of 8.2 MHz. (c) Transmission traces at three values of $E_1$ (labeled in each panel), peak-normalized to emphasize the onset and waveform profiles rather than absolute amplitudes; the red dashed line marks $t = 0$. (d) Onset time $\tau_{\mathrm{o}}$ extracted as a function of $E_1$. $\tau_{\mathrm{o}}$ decreases with increasing $E_1$, indicating a smaller onset time of the oscillation. The red curve is a fit to a piecewise power-law function: $\tau_{\mathrm{o}} = \tau_{\mathrm{off},4}$ for $E_1 \geq E^{''}_{\mathrm{c}}$ and $\tau_{\mathrm{o}} = \tau_{\mathrm{off},4} + a_4(E^{''}_{\mathrm{c}} - E_1)^{\alpha_4}$ for $E_1 < E^{''}_{\mathrm{c}}$, with critical field $E^{''}_{\mathrm{c}} = 19.1$~V/m, plateau value $\tau_{\mathrm{off},4} = 6.4$~$\mu$s, amplitude $a_4 = 18.6$~$\mu$s$\cdot$(V/m)$^{-\alpha_4}$, and critical exponent $\alpha_4 \approx 0.3$ ($R^2 = 0.99$).}
\label{fig4}
\end{figure*}

To further elucidate the parameter space governing the CTC dynamics, we systematically investigated the role of the laser detuning $\Delta_c$ in the oscillatory dynamics. While the RF field amplitude $E$ determines the overall strength of the driving, the laser detuning $\Delta_c$ controls the effective energy mismatch of the Rydberg excitation, thereby tuning the system's approach to the EP in the Liouvillian spectrum. The resulting dynamical map is presented in Fig.~\ref{fig3}(a), where the normalized transmission amplitude is plotted as a function of time and the detuning $\Delta_c$, scanned from $-2\pi \times 7.3$ to $2\pi \times 4.0$~MHz at a fixed RF amplitude $E = 26.5$~V/m. The heatmap reveals four distinct dynamical regimes, labeled A-D. On the red-detuned side (regime C, approximately $\Delta_c < 2\pi \times 0$~MHz), the transient oscillations emerge, but their coherence is rapidly lost, resulting in a short-lived time crystal. On the blue-detuned side (regime D, approximately $\Delta_c \geq 2\pi \times 0$~MHz), the system exhibits a particularly robust CTC with long lifetime.

To quantify this detuning dependence, we extract transmission time traces at detunings $\Delta_c = -2\pi \times 5.5$~MHz (b1), $-2\pi \times 0.5$~MHz (b2), and $2\pi \times 1.8$~MHz (b3), as shown in Fig.~\ref{fig3}(b). At $\Delta_c = -2\pi \times 5.5$~MHz, the oscillations decay within a few hundred microseconds, reflecting a finite real part of the Liouvillian gap. Near resonance ($\Delta_c = -2\pi \times 0.5$~MHz), the oscillations persist for a considerably longer duration before eventual decay, indicating a closing of the dissipative gap. At $\Delta_c = 2\pi \times 1.8$~MHz, the oscillations show no measurable decay within the entire 1.5 ms acquisition window, consistent with the system being deep in the limit-cycle phase where the effective dissipation approaches zero.

Figure~\ref{fig3}(c) summarizes the extracted onset time and lifetime as functions of the detuning $\Delta_c$. The lifetime is fitted with a power-law function: $\tau_{\mathrm{l}}(\Delta_c) = \tau_{\mathrm{off},2} + a_2(\Delta_c/2\pi - \Delta_{\mathrm{min}}/2\pi)^{\alpha_2}$ with $\tau_{\mathrm{off},2} = 272$~$\mu$s, $a_2 = 3.5\times10^{-4}~\mu\mathrm{s}\cdot\mathrm{MHz}^{-\alpha_2}$, $\alpha_2 = 11.7$ ($R^2 = 0.99$), the onset time is fitted to $\tau_{\mathrm{o}}(\Delta_c) = \tau_{\mathrm{off},3} + a_3(\Delta_c/2\pi - \Delta_{\mathrm{min}}/2\pi)^{\alpha_3}$ with $\tau_{\mathrm{off},3} = 83$~$\mu$s, $a_3 = 0.44~\mu\mathrm{s}\cdot\mathrm{MHz}^{-\alpha_3}$, $\alpha_3 = 3.0$ ($R^2 = 0.99$), where $\Delta_{\mathrm{min}}/2\pi = -4.0$~MHz. These results demonstrate that the laser detuning serves as a critical control parameter for the CTC lifetime. The sharp power-law increase on the red-detuned side indicates that the system undergoes a transition from a dissipative steady state to a lossless limit-cycle phase. The blue-detuned configuration brings the real part of the Liouvillian eigenspectrum closest to zero, thereby realizing a nearly dissipationless oscillatory state over the experimentally accessible timescale.

\begin{figure*}
\centering
\includegraphics[width=1\linewidth]{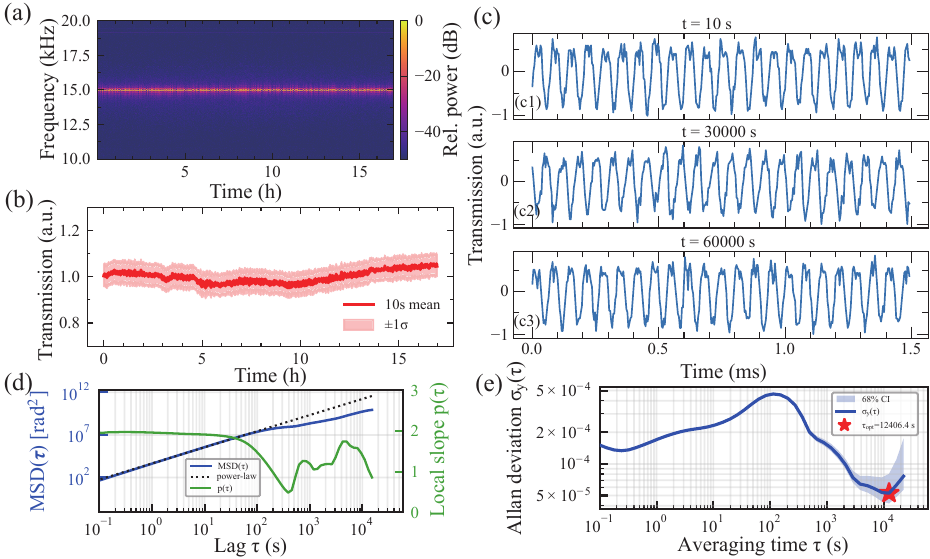}
\caption{\textbf{Measurement of the ultralong-lived CTC.} (a) Short-time Fourier-transform spectrogram of the probe-transmission signal over a 61,020-s analysis window. The oscillation remains centered near \(f_0=15.03~\mathrm{kHz}\), with a maximum frequency offset of \(30.5~\mathrm{Hz}\). The spectral power is normalized to its maximum. (b) Oscillation-amplitude stability over the full lifetime. The red curve shows the 10-s local mean, and the shaded band denotes the local \(\pm1\sigma\) fluctuation. The amplitude remains near its typical value throughout the measurement, with a total variation of about \(10.1\%\). (c) Representative waveforms at \(t=10~\mathrm{s}\), \(30000~\mathrm{s}\), and \(60000~\mathrm{s}\). Each 1.5-ms trace is independently normalized to show the persistence of the microscopic oscillatory waveform. (d) Phase wandering of the time-crystalline oscillation. Mean-squared phase displacement, $\mathrm{MSD}(\tau)$, of the 15 kHz time-crystalline oscillation, evaluated from a 61020 s record. The MSD increases strongly with lag time $\tau$ and follows a superlinear power-law scaling, $\mathrm{MSD}(\tau) \propto \tau^{\beta}$ with $\beta \approx 1.89$, indicating temporally correlated phase wandering rather than simple Brownian phase diffusion. The local logarithmic slope, $p(\tau)=d\log[\mathrm{MSD}(\tau)]/d\log\tau$, plotted on the right axis, approaches $p\simeq1$ only over a limited range of $\tau$, identifying a local diffusive-like regime embedded within long-time drift-dominated dynamics. (e) Overlapping Allan deviation of the fractional frequency. The shaded band denotes the \(68\%\) \(\chi^2\) confidence interval calculated using the Greenhall equivalent degrees of freedom. The Allan deviation reaches \(\sigma_{y,\min}=5.3\times10^{-5}\) at \(\tau_{\mathrm{opt}}=12406~\mathrm{s}\), corresponding to an absolute Allan deviation of approximately \(0.79~\mathrm{Hz}\).
}
\label{fig5}
\end{figure*}

\subsection{Initial-state-dependent CTC}
We next investigate how the transient CTC response depends on the initial state, using a two-stage driving protocol. As shown in Figs \ref{fig4}(a) and (b), the system is first prepared under an RF field $E_1$ (scanned from 0 to 31.3 V/m, $t\leq0$), and then, at $t = 0$, switched to a fixed driving field $E_2 = 31.3$~V/m at the same frequency of 8.2 MHz. The color map of probe transmission versus time and $E_1$ reveals that the oscillation pattern after the switch—its onset, envelope, and phase—varies strongly with the amplitude $E_1$. This demonstrates that the initial state prepared by $E_1$ plays a crucial role in determining the subsequent dynamics, which brings the system closer to the EP in Fig.~\ref{fig1}(b1).

The transmission traces for three specific values of $E_1$ (15.7, 23.5, and 31.3 V/m) are shown in Fig.~\ref{fig4}(c), normalized to highlight waveform evolution rather than absolute amplitude. For $E_1 = 15.7$~V/m, the system starts from a weakly prepared nonoscillatory state, and oscillations build up slowly after $t = 0$, resulting in a long onset time $\tau_{\mathrm{o}}$. At $E_1 = 23.5$~V/m, oscillations appear almost immediately after the switch, indicating that the preparation field has brought the system close to the basin of attraction of the final limit cycle. At \(E_1 = 31.3~\mathrm{V/m}\), oscillations are already present before the switch, indicating preparation in the oscillatory limit-cycle state, and the time-crystalline response is maintained after switching to \(E_2\).

Figure~\ref{fig4}(d) quantifies this trend by plotting the extracted onset time $\tau_{\mathrm{o}}$ as a function of $E_1$. The monotonic decrease of $\tau_{\mathrm{o}}$ with increasing $E_1$ shows that the field $E_1$ prepares the system in a state closer to the eventual oscillatory attractor defined by $E_2$. Physically, the magnitude of the RF field $E_1$ determines the position of the initial state near the EP in the Liouvillian eigenspectrum. Initial states located before the EP require a longer onset time to enter the limit cycle of the time crystal, whereas initial states after the EP already lie in the oscillatory region, substantially reducing the time needed to reach the limit cycle. The results in Fig.~\ref{fig4}(d) illustrate how the onset time gradually decreases and approaches a critical value $E^{''}_{\mathrm{c}}$ as the initial state approaches the EP. Based on our findings, the onset time of oscillations can be systematically shortened by preparing the initial state, offering a practical method for controlling transient dynamics.

\subsection{Ultralong lifetime CTC}
To investigate the lifetime of long-lived oscillations under our optimized experimental conditions, we set the electric field amplitude to \(E=45.6~\mathrm{V/m}\) and the laser detuning to \(\Delta_c=0~\mathrm{MHz}\). These conditions place the system in the regime where the transient decay observed in the short-time scans is strongly suppressed, which corresponds to the near-zero real part of the Liouvillian eigenvalues. Under these conditions, we measure the maximum lifetime that CTCs can attain in our experimental configuration. Observing ultralong lifetime in CTCs imposes stringent demands on the long-term stability of the experimental apparatus. In particular, the excitation lasers must remain frequency-locked with high precision over the entire measurement duration, as detailed in the Methods section. Unlike the triggered measurements used in Figs.~\ref{fig2}--\ref{fig4}, the long-time record in Fig.~\ref{fig5} was acquired in continuous streaming mode. We analyze the stable oscillatory portion of the record from \(t=10~\mathrm{s}\) to \(t=61030~\mathrm{s}\), corresponding to a continuous observation window of \(61020~\mathrm{s}\), or \(16.95~\mathrm{h}\).

We recorded both the Fourier spectrogram and the time-domain oscillation amplitude, as shown in Fig.~\ref{fig5}(a,b). The Fourier spectrogram in Fig.~\ref{fig5}(a) remains centered near \(f_0=15.03~\mathrm{kHz}\) throughout the full \(16.95~\mathrm{h}\) analysis window, with a maximum frequency fluctuation of \(30.5~\mathrm{Hz}\), indicating a stable and persistent oscillatory mode. The time-domain oscillation amplitude in Fig.~\ref{fig5}(b) remains nearly constant, with no systematic collapse over the measurement window. To illustrate the temporal persistence of the CTC, Fig.~\ref{fig5}(c) presents three independently normalized waveforms extracted at \(t=10~\mathrm{s}\), \(30000~\mathrm{s}\) and \(60000~\mathrm{s}\). The stable frequency and amplitude indicate that the system remains close to a robust limit-cycle orbit throughout the measurement.

In a dissipative limit cycle, radial fluctuations are restored toward the attractor. Residual technical noise and slow environmental perturbations drive phase wandering without implying decay of the time-crystalline order. We quantify this angular motion using the mean-squared phase displacement $\mathrm{MSD}(\tau)=\langle[\psi(t+\tau)-\psi(t)]^2\rangle$, where \(\psi(t)\) is the unwrapped phase. As shown in Fig.~\ref{fig5}(d), the MSD grows over several orders of magnitude and approximately follows \(\mathrm{MSD}(\tau)\propto\tau^{\beta}\), with \(\beta\approx1.89\). The local logarithmic slope, $p(\tau)=d\log[\mathrm{MSD}(\tau)]/{d\log\tau}$, approaches \(p\approx1\) only over a limited range of lag times. Thus the phase dynamics are not described by simple random diffusion over the full record, but instead reflect correlated, slow-drift-induced wandering of the angular coordinate.

Furthermore, we quantify the stability of the oscillation frequency using the overlapping Allan deviation [Fig.~\ref{fig5}(e)]. For the fractional frequency \(y(t)=[f(t)-f_0]/f_0\), the Allan variance is $\sigma_y^2(\tau)=\frac{1}{2}\langle(\bar y_{k+1}-\bar y_k)^2\rangle$, where \(\bar y_k\) is the average fractional frequency over the \(k\)-th interval of duration \(\tau\). This quantity compares adjacent frequency averages and is therefore appropriate for oscillators affected by slow drift. The Allan deviation reaches a minimum of \(\sigma_{y,\min}=5.3\times10^{-5}\) at \(\tau_{\mathrm{opt}}=12406~\mathrm{s}\), corresponding to an absolute Allan deviation of approximately \(0.79~\mathrm{Hz}\). The non-monotonic dependence on averaging time reflects frequency instabilities on different timescales: short-to-intermediate-time fluctuations increase \(\sigma_y(\tau)\) over \(\tau \sim 10^{-1}\)–\(10^{2}~\mathrm{s}\), whereas averaging suppresses these fluctuations over \(\tau \sim 10^{2}\)–\(10^{4}~\mathrm{s}\), leading to a broad minimum near \(10^4~\mathrm{s}\). The slight upturn at the longest averaging times suggests the onset of residual slow drift.

\section*{Discussions}
The reported lifetime exceeding 16.95 hours in the Rydberg atom system represents a dramatic advancement over previous CTC demonstrations, which have typically been limited to milliseconds in atomic systems and approximately 40 minutes in electron-nuclear spin systems. The suppression of transverse relaxation to such an extreme degree implies that the Liouvillian spectral gap remains effectively closed over macroscopic times, offering a concrete experimental realization of a stable limit-cycle phase in the thermodynamic limit. 

With this ultralong lifetime, several previously inaccessible directions become experimentally feasible. First, CTCs may serve as promising quantum sensors for external fields, where the persistent oscillatory phase enables long integration times and stable frequency readout over extended durations. Second, the robustness of the time-crystalline order against decay opens the possibility of using CTCs as continuous-time quantum memory elements, where information is encoded in the phase or amplitude of the limit-cycle oscillation and can be preserved for arbitrarily long durations under steady pumping. Third, the ability to maintain lifetime over 16.95 hours allows for systematic studies of fundamental questions, such as whether small but finite many-body interactions or residual thermal effects eventually destroy the time crystal beyond the experimentally accessible window, and whether quantum fluctuations induce a finite but extremely long lifetime even when the mean-field Liouvillian gap closes. 

The strong, long-range Rydberg interactions play an important role in determining the CTC lifetime. On one hand, the van der Waals interactions introduce strong nonlinearity for limit-cycle formation, enabling the system to escape the stationary steady state and enter an oscillatory phase. On the other hand, these long-range interactions can be a source of dephasing and thermalization if not properly balanced by dissipation. In our experiment, the dissipative environment—engineered via controlled optical pumping and RF dressing—is carefully tuned to counteract interaction-induced decay, thereby stabilizing the limit-cycle attractor. Our results thus provide a benchmark for understanding the fundamental limits of dissipative time crystals and offer a scalable path toward their potential application in frequency reference in quantum sensing \cite{cabot2024continuous,o2025quantum} and memory elements in quantum computers \cite{makinen2025continuous,casey2025time}. 

\section*{Summary}
In this work, we experimentally realize a CTC in a driven-dissipative Rydberg atom ensemble with an ultralong lifetime exceeding 16.95 hours. By engineering strong long-range Rydberg interactions and a controlled dissipative environment, we stabilize the limit-cycle attractor and suppress the decay and heating effects that typically destroy time-crystalline order. The control of the Liouvillian spectrum is vital to realizing an ultralong lifetime CTC. In particular, by systematically tuning the system parameters, we are able to close the gap in the real part of the Liouvillian eigenvalues and bring it to near-zero, which underlies the suppression of dissipation and the emergence of a persistent oscillatory phase. Through systematic optimization of the electric field amplitude, laser detuning, and initial state preparation, we realize an oscillatory phase whose lifetime is orders of magnitude longer than in previous CTC demonstrations. Our experimental results show good agreement with theoretical calculations based on the Liouvillian spectral analysis and the mean-field master equation, thereby validating the predicted emergence of a long-lived limit-cycle phase. This work establishes a robust platform for exploring autonomous nonequilibrium quantum phases.  

\section*{Methods}
\subsection*{Experimental setup}
The experimental apparatus for the preparation and measurement of the CTC was constructed based on a three-photon Rydberg excitation scheme in Caesium atoms. The excitation pathway involves four atomic states, \(\left|6S_{1/2}\right\rangle\), \(\left|6P_{3/2}\right\rangle\), \(\left|7S_{1/2}\right\rangle\), and \(\left|49P_{3/2}\right\rangle\). An 852 nm probe laser drives the transition from the ground state \(\left|6S_{1/2}\right\rangle\) to the first intermediate state \(\left|6P_{3/2}\right\rangle\). A 1470 nm dressing laser further excites the atoms to the second intermediate state \(\left|7S_{1/2}\right\rangle\), and a 780 nm coupling laser subsequently drives the transition to the Rydberg state \(\left|49P_{3/2}\right\rangle\). A pair of RF electrodes was mounted on both sides of the atomic vapor cell to apply an external RF field to the interaction region. When the system parameters were tuned close to the limit-cycle regime, periodic oscillations associated with the CTC phase transition were observed in the optical response of the atomic ensemble.

The 852 nm probe laser and the 1470 nm dressing laser were generated by diode laser systems (TOPTICA DL pro and TOPTICA DL 100, respectively). The 780 nm coupling laser was amplified using a tapered amplifier system (TOPTICA TA pro) before entering the optical setup to provide the power required for Rydberg excitation. The powers of the probe, dressing, and coupling lasers were 214 $\mu$W, 26 mW, and 1.5 W, respectively, with corresponding \(1/e^2\) beam waist radii of approximately 270 \(\mu\)m, 500 \(\mu\)m, and 340 \(\mu\)m. The corresponding Rabi frequencies were \(\Omega_p = 2\pi \times \mathrm{43}\) MHz, \(\Omega_d = 2\pi \times \mathrm{384}\) MHz, and \(\Omega_c = 2\pi \times \mathrm{10.4}\) MHz, respectively.

The three laser beams were combined, collimated, and spatially overlapped using mirrors, dichroic mirrors, and collimating optics, forming a common interaction region inside the Caesium vapor cell. The transmitted probe beam was detected by a photodetector (PDB450A-DC) as the primary experimental observable. For the short-time measurements in Figs.~\ref{fig2}--\ref{fig4}, the electrical output was recorded by an oscilloscope (RIGOL DHO1204U), together with the RF driving signal and the trigger signal, to ensure temporal correspondence between different channels. For the long-time continuous measurement in Fig.~\ref{fig5}, the photodetector output was recorded using a PicoScope 2208B operated in streaming mode. Additional details of the experimental setup are provided in Supplementary Fig.~S1 and the Supplementary Information.

\subsection*{External RF field}
The external RF field was applied using a pair of parallel electrode plates mounted on both sides of the vapor cell. The electrodes had a diameter of 120 mm, a thickness of 3 mm, and a separation of 40 mm. The RF signal was generated by an arbitrary function generator (RIGOL DG902 PRO) and delivered to the electrode structure through connecting cables. The output frequency, amplitude, trigger mode, and temporal waveform sequence of the function generator were configured by a computer. In the experiment, the RF drive was operated at a fixed frequency of \(8.2~\mathrm{MHz}\), while the atomic system was driven by varying either the RF amplitude or the temporal structure of the RF signal.

The specially designed temporal RF sequence was implemented using the dual-channel output of the arbitrary function generator. The two channels were trigger-controlled to form a composite RF sequence in the time domain, consisting of a preparation segment followed by an evolution segment with a fixed amplitude. The synchronization trigger generated by the function generator was delivered to the data acquisition device to define the correspondence between the RF switching time and the time origin of the recorded experimental signal.

\subsection*{Long-term stability of lasers}
To support extended CTC preparation and measurement, the 852 nm probe laser was stabilized in frequency using saturated absorption spectroscopy (SAS), the 1470 nm dressing laser was locked using a two-photon electromagnetically induced transparency (EIT) spectrum, and the 780 nm coupling laser was stabilized using a three-photon EIT spectrum. Before data acquisition, the laser systems, tapered amplifier and associated optical components were allowed to warm up for 1 h to reach stable operating conditions. During a continuous 16.95 h measurement, all three lasers remained locked. The detected signal was recorded in real time as temporal waveforms to track the measured response over the full acquisition period. 

\subsection*{Master Equations}
Beyond computing the Liouvillian spectrum of the system to account for the formation of the time crystal, we also calculate the dynamical evolution of the system via the Lindblad master equation approach. Owing to the thermal motion of the atoms, correlations between atoms can be neglected, allowing us to adopt the mean-field approximation. From the Hamiltonian of the system, we obtain the master equation

\allowdisplaybreaks

\begin{subequations}
\begin{align}
\dot{\rho}_{gR} =& \frac{i}{2}\Big(i\gamma\rho_{gR}+2\Delta\rho_{gR}+(\rho_{gg}-\rho_{RR})\Omega \notag\\
&-\rho_{R_1R}\Omega_{1}-\rho_{R_2R}\Omega_{2}\Big), \\[4pt]
\dot{\rho}_{gR_1} =& \frac{i}{2}\Big(i\gamma_{1}\rho_{gR_1}+2\Delta_{1}\rho_{gR_1}-\rho_{RR_1}\Omega \notag\\
&+(\rho_{gg}-\rho_{R_1R_1})\Omega_{1}-\rho_{R_2R_1}\Omega_{2}\Big), \\[4pt]
\dot{\rho}_{gR_2} =& \frac{i}{2}\Big(
i\gamma_{2}\rho_{gR_2}
+2\Delta_{2}\rho_{gR_2}
-\rho_{RR_2}\Omega
-\rho_{R_1R_2}\Omega_{1} \notag\\
&+(\rho_{gg}-\rho_{R_2R_2})\Omega_{2}\Big), \\[4pt]
\dot{\rho}_{RR} =& -\tfrac{i}{2}\Omega\,\rho_{gR} + \tfrac{i}{2}\Omega\,\rho_{Rg} - \gamma\,\rho_{RR}, \\[4pt]
\dot{\rho}_{RR_1} =& -\frac{i}{2}\Big(
(-i(\gamma+\gamma_{1})+2\Delta-2\Delta_{1})\rho_{RR_1} \notag\\
&+\rho_{gR_1}\Omega
-\rho_{Rg}\Omega_{1}\Big), \\[4pt]
\dot{\rho}_{RR_2} =&-\frac{i}{2}\Big(
(-i(\gamma+\gamma_{2})+2\Delta-2\Delta_{2})\rho_{RR_2} \notag\\
&+\rho_{gR_2}\Omega
-\rho_{Rg}\Omega_{2}\Big), \\[4pt]
\dot{\rho}_{R_1 R_1} =& -\gamma_{1}\rho_{R_1R_1}
-\frac{i}{2}
(\rho_{gR_1}-\rho_{R_1g})\Omega_{1}, \\[4pt]
\dot{\rho}_{R_1 R_2} =& -\frac{i}{2}\Big(
(-i(\gamma_{1}+\gamma_{2})+2\Delta_{1}-2\Delta_{2})\rho_{R_1R_2} \notag\\
&+\rho_{gR_2}\Omega_{1}
-\rho_{R_1g}\Omega_{2}\Big), \\[4pt]
\dot{\rho}_{R_2 R_2} =& -\gamma_{2}\rho_{R_2R_2}
-\frac{i}{2}
(\rho_{gR_2}-\rho_{R_2g})\Omega_{2}.
\end{align}
\end{subequations}

The other expressions satisfy $\rho_{Rg}=\rho_{gR}^\dagger$, $\rho_{R_1 g}=\rho_{gR_1}^\dagger$, $\rho_{R_2 g}=\rho_{gR_2}^\dagger$, $\rho_{R_1 R}=\rho_{RR_1}^\dagger$, $\rho_{R_2 R}=\rho_{RR_2}^\dagger$, $\rho_{R_2 R_1}=\rho_{R_1 R_2}^\dagger$, and the Rydberg population is normalized with $1 = \rho_{gg} + \rho_{R_1R_1} + \rho_{RR} + \rho_{R_2 R_2}$. Within the mean-field approximation, the detuning $\Delta$, $\Delta_1$ and $\Delta_2$ in the master equation are modified to include a correction arising from the interactions between Rydberg atoms. By numerically solving the master equation, we obtain the dynamical evolution of the system, as shown in Fig.~\ref{fig1}(c).

\section*{Acknowledgements}
We acknowledge funding from the National Natural Science Foundation of China (Grant Nos. T2495253, 62435018), the National Key R and D Program of China (Grant No. 2022YFA1404002).

\section*{Data Availability}
All experimental data used in this study are available from the corresponding author upon request.

\section*{Author contributions statement}
D.-S.D. conceived the idea and supported this research. Q.-F.W., T.-Y.H. and D.-Y.Z. conducted the physical experiments. Y.-J.W. conducted the theoretical calculations. The manuscript was written by D.-S.D., B.L., Q.-F.W., T.-Y.H., D.-Y.Z., and Y.-J.W. All authors contributed to discussions regarding the results and the analysis contained in the manuscript.

\section*{Competing interests}
The authors declare no competing interests.

\balance
\bibliography{ref}

\end{document}